\documentclass[12pt]{article}
\begin{document}
\baselineskip=0.20in
\vspace{20mm}
\baselineskip=0.30in
\begin{center}

{\large \bf Approximate Spin and Pseudospin Solutions of the Dirac equation with Rosen-Morse Potential including a Coulomb Tensor Interaction}
\vspace{10mm}

{  K. J. Oyewumi\footnote{International Chair in Mathematical Physics and Applications (ICMPA-UNESCO Chair), \\University of Abomey-Calavi, $072$ BP $50$ Cotonou, Republic of Benin. E-mail: kjoyewumi66@unilorin.edu.ng or~ mjpysics@yahoo.com}\\
Theoretical Physics Section, Department of Physics,\\ University of Ilorin,   P. M. B. 1515, Ilorin, Kwara state, Nigeria.}
 \vspace{5mm}

\baselineskip=0.20in

\vspace{4mm}

\end{center}

\noindent
\begin{abstract}
By applying the Pekeris-type approximation to deal with the (pseudo or) centrifugal term, the spin and pseudospin symmetry solutions of the Dirac equation for the Rosen-Morse potential including a Coulomb-like tensor potential with arbitrary spin-orbit coupling quantum number $\kappa$  are obtained by standard method. It has been shown from the numerical results that the degeneracies between spin and pseudospin state doublets are removed by the tensor interaction. Special case of this potential, that is, the spin and pseudospin solutions of the Dirac equation with  the modified P\"{o}schl-Teller potential including a tensor interaction is also considered. The results obtained in this case show that the tensor interaction removes the degeneracies between the members of doublets states in spin and pseudospin symmetries.
\end{abstract}

 \vspace{2mm}

 \vspace{2mm}

\section{Introduction}
For over $40$ years ago, the idea about the pseudospin symmetry (PSS) and spin symmetry (SS) concepts have been introduced in nuclear theory by Arima et al. (1969) and Hecht and Alder (1969) \cite{ArE69, HeA69}. The PSS refers to a quasi degeneracy of single nucleon doublets with nonrelativistic quantum number $(n, \ell, j = \ell + \frac{1}{2})$ and $(n - 1, \ell + 2, j = \ell + \frac{3}{2})$, where $n, \ell~ \mbox{and}~j$ are the radial, the orbital and the total angular quantum numbers, respectively \cite{ArE69, HeA69, PaE01}. The total angular momentum is $j = \overline{\ell} + \overline{s}$, where $\overline{\ell} = \ell + 1$ is the pseudo-angular momentum and $\overline{s}$ is the pseudospin angular momentum \cite{IkS10}. 

Pseudospin doublets with pseudo-orbital angular momentum $\overline{\ell} = \ell + 1$ and pseudospin $\overline{s} = \frac{1}{2}$ quantum numbers are the single-nucleon states that have energy values close to each other \cite{AyS101}. For example, $(3s_{1/2}, 2d_{2/3})$ and $(3\overline{p}_{1/2}, 2\overline{d}_{2/3})$ can be considered as pseudospin doublets.  These concepts have been used in describing deformed nuclei \cite{BoE82}, superdeformation \cite{DuE87}, magnetic moment \cite{TrE94}, meson theory \cite{PaE04}, as well as establishing an effective shell-model coupling scheme \cite{TrE95}.

The introduction of these concepts has given room for considerable research efforts in nuclear theory as well as quantum theory. In his investigations, Ginocchio (1997, 1999, 2004, 2005a, 2005b) deduced that, the Dirac Hamiltonian with the scalar and vector harmonic oscillator potentials for the case $V(r) = S(r)$ possesses a spin symmetry as well as $U(3)$ symmetry, while the Dirac Hamiltonian for the case $V(r) = S(r)$ possesses a pseudospin symmetry as well as a pseudo-$U(3)$ symmetry [11 - 15 ]. 
In addition, Meng et al. (1998) and Zhou et al. (2003) found that PSS is exact under the condition $V(r) + S(r) = C_{ps}$, while SS is exact under the condition $V(r) - S(r) = C_{s}$  [11 - 15 ]. 
For comprehensive reviews, see Ginocchio (1997) and (2005b) \cite{Gin97, Gin052}.

Under the condition of PSS and SS concepts, the Dirac Hamiltonian for some exactly solvable potential models have been solved for any $\ell$ [12, 14, 15, 19 - 33]. 
Some of these potential models can not be solved exactly for any $\ell \neq 0$, in which the Pekeris-type approximation can be used to obtain the approximate solutions for such models [5, 34 - 48]. 
In solving these problems, various methods have been adopted, these methods include: Nikiforov-Uvarov method, AIM, SUSYQM and  standard method as well. 

In their studies, Moshinsky and Szczepaniak (1989) and Kukulin et al. (1991) introduced tensor potential $U(r)$ into the Dirac equation, by replacing $\vec{p}$ with $\vec{p} - im \omega \beta . \hat{r}U(r)$ (that is, $\vec{p} \rightarrow \vec{p} - im \omega \beta . \hat{r}U(r)$) \cite{MoS89, KuE91}. The inclusion of tensor interaction or coupling was first introduced by  It$\hat{o}$ et al. (1969) \cite{ItE69} and this has been revived by Moshinsky and Szczepaniak (1989) \cite{MoS89}. 

Akcay (2007) studied the Dirac equation with a tensor potential which contains a term linear in $r$ and a Coulomb-like term, the eigenstates and eigenvalues are obtained analytically \cite{Akc07}. The Dirac equation with the scalar and vector quadratic potentials and Coulomb-like tensor potential with the PSS and SS have been solved exactly, and the results discussed \cite{AkT09, Akc09}. 
Tensor couplings or interactions have been used successfully in the studies of nuclear properties and applications [4, 5, 19 - 27, 31 - 68]. 

In some of the studies of the PSS and SS, only few of the known model potentials are exactly solvable for any $\ell$, while approximate solutions can be obtained for any $\ell$ for some potentials. Exact SS and PSS solutions of the Dirac equations of a linear term, quadratic, Mie-type, pseudoharmonic, Coulomb and harmonic potentials coupled with Coulomb-like tensor interactions have been obtained by using different methods \cite{ AkT09, Akc07, Akc09, AyS09, HamE101,  HamE102, ZaE10, ZaE11,  AyS104,  Typ08}.
 
In order to obtain the approximate solutions for some model potentials that can not be exactly solved, the Pekeris-type approximation needs to be employed \cite{Pek34, GrA76}. In this manner, the PSS solutions of the Dirac equation with the Eckart potential including a Coulomb-like tensor potential with arbitrary spin-orbit coupling quantum number $\kappa$ are obtained by using the Nikiforov-Uvarov method \cite{HamE11}. The Dirac equation is solved approximately for the Woods-Saxon potential and a tensor potential with the arbitrary spin-orbit coupling quantum number $\kappa$ under pseudospin and spin symmetry via the standard method \cite{AyS101}. 

Ikhdair and Sever (2010) obtained the approximate solutions of the Dirac equation with the Hulth$\acute{e}$n potential including Coulomb-like tensor for any $\kappa$ under the SS and PSS limits \cite{IkS10}. Pseudospin, supersymmetry and the shell structure of atomic nuclei have been discussed by Typel (2008), he found that the strength of the pseudospin splitting depends on isovector-dependent and tensor contributions to the effective nuclear interaction \cite{Typ08}. In addition, by using the SUSYQM method, Hassanabadi et al. (2011) and (2012) obtained the SS and PSS solutions of the Dirac equation with a Coulomb tensor interaction for the scalar and vector hyperbolic and Tietz  potentials, respectively    \cite{HasE11, HasE121}.

The SS solution of the Dirac equation with position-dependent mass for $q$-parameter modified P\"{o}schl-Teller and Coulomb-like tensor potential has been obtained \cite{EsM112}. Approximate solutions of the Dirac equation for the generalized P\"{o}schl-Teller scalar and vector potentials and a Coulomb tensor interaction via the Nikiforov-Uvarov method have been obtained by Hassanabadi et al. (2012) \cite{HasE122}. Their results find application in both nuclear and hadron physics and provide the more general solutions to the previous works of Agboola (2011), Wei and Dong (2009) \cite{Agb11, WeD09}. 

Hamzavi et al. (2012a) used the Nikiforov-Uvarov method to study the relativistic Morse potential and tensor interaction, and found that tensor interaction removes degeneracies between each pairs of pseudopin and spin doublets \cite{HamE121}. Recently, Maghsoodi et al. (2012) used the SUSYQM approach to obtain the spectrum of the Dirac equation under the Deng- Fan scalar and vector potentials and a Coulomb tensor interaction \cite{MaE12}. From their results, it was found that tensor interactions remove all degeneracies between the two states in pseudospin and spin doublets. Also, they found that the energy differences between the two states in pseudospin and spin doublets increase with increasing $H$.

Furthermore, in the framework of the spin and pseudospin symmetry, Hamzavi et al.(2012b) solved the Dirac equation for the inversely quadratic Yukawa potential including a Coulomb-like tensor potential with arbitrary spin-orbit coupling quantum number $\kappa$ by using the Nikiforov-Uvarov method \cite{HamE122}. The numerical results obtained show that the Coulomb-like tensor interaction removes degeneracies between spin and pseudospin state doublets.

In this study, we consider the Rosen-Morse potential, due to its applications in atomic, chemical and molecular Physics \cite{RoM32}. The Rosen-Morse potential is given as
\begin{equation}
V(r)= -V_{1}  \textrm{sech}^{2}{\alpha r} + V_{2}\tanh{\alpha r},
\label{t1}
\end{equation}
where $V_{1}$ and $V_{2}$ are the depth of the potential and $\alpha$ is the range of the potential, respectively. This potential model has been used in so many investigations in theoretical physics [39, 74 - 83]. 

In this letter, we solve the Dirac equation with the scalar and vector Rosen-Morse potentials including a Coulomb tensor interaction under the PSS and SS limits, this is can be achieved by using the Pekeris-type approximation scheme for the (pseudo or) centrifugal term via the standard method. 

This letter is arranged as follows: in Section $2$, we discussed the Dirac equation with tensor potential under the PSS and SS limits. In the presence of spin and pseudospin symmetry limits, we obtain the bound state solutions of the Dirac equation with the Rosen-Morse potential including a tensor interaction in Section $3$.  Section $4$ contains the SS and PSS solutions of a special case of the Dirac equation with the Rosen-Morse potential (modified P\"{o}schl-Teller) including a tensor interaction. The relevant concluding remarks are given in Section $5$.

\section{The Dirac Equation Including a Tensor Interaction}
In spherical coordinates, the Dirac equation for fermionic massive spin-$\frac{1}{2}$ particles moving in attractive scalar $S(r)$, repulsive vector potential and tensor $U(r)$ potential is given as ($\hbar = c = 1$) [4, 5, 19, 20, 21, 23 - 27, 31 - 35, 49 - 84].  
\begin{equation}
\left[ { \vec{\alpha} }. { \vec{P}} + \beta [M + S(\vec{r}) ] - i \beta \vec{\alpha}~.~ \hat{r} U(r)\right]\psi_{n\kappa}(\vec{r}) =  [E - V(\vec{r})]\psi_{n\kappa}(\vec{r}),
\label{t2}
\end{equation}
where $E$ is the relativistic energy of the system, $M$ is the mass of a particle, $ \vec{P} = -i \hbar \nabla$ is the momentum operator. $\vec{\alpha}$ and $\beta$ are $4 \times 4$ Dirac matrices, given as
\begin{equation}
 ~~{\vec{ \alpha}}= \left(\matrix {
0 & \sigma_{i} \cr
\sigma_{i} & 0\cr
}\right), ~~
 \beta= \left(\matrix {
I & 0 \cr
0 & -I \cr
}\right), 
\label{t3}
\end{equation}
where $I$ is the $2 \times 2$ identity matrix and $\sigma_{i}(i = 1, 2, 3)$  are the vector Pauli matrices.

The spinor wave functions can be written using the Pauli-Dirac representation as \cite{Ikh10, Ikh11, IkE11,   WeD101, WeD102, WeD103, WeD104,   WeD09, OyA10, Gre00}:
\begin{equation}
\psi_{n\kappa}(\vec{r})=  
\frac{1}{r}\left[\matrix {
F_{n\kappa}(r) ~Y^{\ell}_{jm}(\theta, \phi)\cr
iG_{n\kappa}(r)~Y^{\overline{\ell}}_{jm}(\theta, \phi)\cr
}\right],~ \kappa = \pm(j + \frac{1}{2}), 
\label{t4}
\end{equation}
where $F_{n\kappa}(r)$ and $G_{n\kappa}(r)$ are the radial wave functions of the upper and lower spinors components, respectively. $Y^{\ell}_{jm}(\theta, \phi)$ and $Y^{\overline{\ell}}_{jm}$ are the spherical harmonic functions coupled to the total angular momentum $j$ and its projection $m$ on the $z-$axis. The orbital and pseudo-orbital angular momentum quantum numbers for SS ($\ell$) and PSS ($\overline{\ell}$) refer to the upper ($F_{n\kappa}(r)$) and lower ($G_{n\kappa}(r)$) spinor components, respectively, for which $\ell(\ell + 1) = \kappa(\kappa + 1)$ and $\overline{\ell}(\overline{\ell} + 1) = \kappa(\kappa - 1)$. For comprehensive reviews, see Ginocchio (1997), (2005b) and Ikhdair et al. (2011)  \cite{Gin97, Gin052, IkE11}.

By using  the following relations (Bjorken and Drell 1964) \cite{BjD64}: 
\begin{equation}
(\vec{\sigma}~ .~ \vec{A})(\vec{\sigma}~ .~ \vec{B}) = \vec{A}~ .~ \vec{B} + i  \vec{\sigma}~ .~ (\vec{A}\times \vec{B}), 
\label{t5}
\end{equation}
\begin{equation}
(\vec{\sigma}~ .~ \vec{P}) = \vec{\sigma}~ .~ \hat{r} \left( \hat{r}~ .~\vec{\sigma} + i \frac{\vec{\sigma}~.~ \vec{L}}{r} \right), 
\label{t6}
\end{equation}
with the following properties being satisfied:
\begin{eqnarray}
&(\vec{\sigma}~ .~ \vec{L})Y_{jm}^{\overline{\ell}}(\theta, \phi) = (\kappa - 1)Y_{jm}^{\overline{\ell}}(\theta, \phi), \nonumber \\
&(\vec{\sigma}~ .~ \vec{L})Y_{jm}^{\ell}(\theta, \phi) = - (\kappa - 1)Y_{jm}^{\ell}(\theta, \phi),\nonumber \\
&(\vec{\sigma}~ .~ \hat{r})Y_{jm}^{\overline{\ell}}(\theta, \phi) = - Y_{jm}^{\ell}(\theta, \phi),
\nonumber \\
&(\vec{\sigma}~ .~ \hat{r})Y_{jm}^{\ell}(\theta, \phi) = - Y_{jm}^{\overline{\ell}}(\theta, \phi).
\label{t7}
\end{eqnarray}
By substituting equation (\ref{t4}) into equation (\ref{t2}), the following two radial coupled differential equations for the upper and the lower component spinors $F_{n\kappa}(r)$ and $G_{n\kappa}(r)$ are obtained as:
\begin{equation}
\left[  \frac{d}{dr} + \frac{\kappa}{r} - U(r)\right]F_{n \kappa}(r) =  
\left[M + E_{n \kappa} - \Delta(r)\right] G_{n \kappa}(r)
\label{t8},
\end{equation}
\begin{equation}
\left[  \frac{d}{dr} - \frac{\kappa}{r} + U(r)\right]G_{n \kappa}(r) =  
\left[M - E_{n \kappa} + \Sigma(r)\right] F_{n \kappa}(r)
\label{t9},
\end{equation}
where $\Delta(r) = V(r) - S(r)$ and $\Sigma(r) = V(r) + S(r)$ are the difference and sum potentials, respectively.

The following two Schr\"{o}dinger-like differential equations for the radial upper and lower component spinors are obtained, respectively, by eliminating $F_{n\kappa}(r)$ and $G_{n\kappa}(r)$ from equations (\ref{t8}) and (\ref{t9}) to have the following equations:
\begin{eqnarray}
&\displaystyle{\left\{ \frac{d^{2}}{dr^{2}} - \frac{\kappa(\kappa + 1)}{r^{2}} +\frac{2\kappa}{r}U(r) - \frac{dU(r)}{dr} - U^{2}(r) + \frac{\frac{d \Delta(r)}{dr} }{\left[M + E_{n \kappa} - \Delta(r)\right]}\left(\frac{d}{dr} + \frac{\kappa}{r} - U(r) \right)  \right\}F_{n \kappa}(r)} \nonumber \\
& \displaystyle{ -   \left\{\left[M + E_{n \kappa} - \Delta(r)\right] \left[M - E_{n \kappa} + \Sigma(r)\right] \right\} F_{n \kappa}(r)} = 0
\label{t10},
\end{eqnarray}
\begin{eqnarray}
&\displaystyle{\left\{ \frac{d^{2}}{dr^{2}} - \frac{\kappa(\kappa - 1)}{r^{2}} +\frac{2\kappa}{r}U(r) + \frac{dU(r)}{dr} - U^{2}(r) + \frac{\frac{d \Sigma(r)}{dr} }{\left[M - E_{n \kappa} + \Sigma(r)\right]}\left(\frac{d}{dr} - \frac{\kappa}{r} + U(r) \right)  \right\}G_{n \kappa}(r)} \nonumber \\
& \displaystyle{ -   \left\{\left[M + E_{n \kappa} - \Delta(r)\right] \left[M - E_{n \kappa} + \Sigma(r)\right] \right\} G_{n \kappa}(r)} = 0
\label{t11},
\end{eqnarray}
where $\kappa(\kappa - 1) = \overline{\ell}(\overline{\ell} + 1)$ and $\kappa(\kappa + 1) = \ell(\ell + 1)$.

In this case,  a Coulomb-like potential as a tensor potential is added to the Rosen-Morse potential. This Coulomb-like potential is given as,
\begin{equation}
U(r) = -\frac{H}{r}, ~~~~~~~~~~~~~~~~~~H = \frac{Z_{a}Z_{b}e^{2}}{4\pi\epsilon_{0}},~~~~~~~~~~~~~~~r\geq R_{c}
\label{t12},
\end{equation}
where $R_{c} = 7.78\mbox{fm}$ is the Coulomb radius, $Z_{a}$ and $Z_{b}$ respectively, are the charges of the projectile $a$ and the target nuclei $b$ [4, 31, 32, 34, 64 - 68, 86]. 

\section{Solutions of the Dirac equation with the Rosen-Morse potential and a tensor interaction}

\subsection{Spin Symmetric Bound State Solutions}
The exact spin symmetry occurs in the Dirac equation when $\frac{d\Delta(r)}{dr} = 0$ or $\Sigma(r) = C_{s} = \mbox{constant}$. Under this symmetry condition, the equation for the upper radial component spinor $F_{n\kappa}$ for the Rosen-Morse potential plus a Coulomb-like tensor potential becomes
\begin{equation}
\left\{ \frac{d^{2}}{dr^{2}} - \frac{\kappa(\kappa + 1)}{r^{2}} -\frac{2\kappa}{r^{2}}H - \frac{H}{r^{2}} -  \frac{H^{2}}{r^{2}}  - \left[M + E_{n \kappa} - C_{s}\right] \left[M - E_{n \kappa} + \Sigma(r)\right] \right\} F_{n \kappa}(r) = 0
\label{t13},
\end{equation}
where $ \kappa = \left\{
\begin{array}{ll}
\ell, & \mbox{for}\; \kappa < 0  \\
- (\ell + 1), & \mbox{for}\; \kappa >0~~,
\end{array}  \right. $ 
respectively. 

It is known that only the $s$-wave ($\kappa = 0, -1$ or $\kappa = 0, 1$) solutions for the Rosen-Morse potential can be obtained exactly \cite{OyA10, Oye12, IkS12}, an improved new approximation (Pekeris-type) in dealing with the spin-orbit (or pseudo) centrifugal term $\frac{\kappa(\kappa + 1)}{r^{2}}$ (or $\frac{\kappa(\kappa - 1)}{r^{2}}$) to obtain the approximate solutions of this potential is required. The Pekeris approximation introduced by Pekeris (1932) and other form of this approximation for short-range potential was also proposed by Greene and Aldrich (1976) to the centrifugal term ($1/r^{2}$) \cite{Pek34, GrA76}. The Pekeris-type approximation has been successfully applied to the spin-orbit (or pseudo) centrifugal term $\frac{\kappa(\kappa + 1)}{r^{2}}$ (or $\frac{\kappa(\kappa - 1)}{r^{2}}$) by Oyewumi (2012) and references therein \cite{Oye12}.

The centrifugal (or pseudo centrifugal)~ approximation introduced by Lu (2005), Ikhdair (2010) and Oyewumi (2012) for values of $\kappa$ (that are not large and small amplitude of the vibrations about the minimum point $r = r_{0}$) is adopted as follows \cite{Ikh10, Oye12, IkS12, Lu05}:
\begin{equation}
\frac{1}{r^{2}} \approx  \frac{1}{r_{0}^{2}} \left[ c_{0} + c_{1} \left(\frac{-e^{-2\alpha r}}{1 + e^{-2\alpha r}}\right) + c_{2}\left(\frac{-e^{-2\alpha r}}{1 + e^{-2\alpha r}}\right)^{2} \right]
\label{t14},
\end{equation}
\begin{eqnarray}
&C_{0} &= 1 - \left( \frac{1 + e^{-2\alpha r_{0}}}{2\alpha r_{0}}\right)^{2} \left(\frac{8 \alpha r_{0}}{1 + e^{-2\alpha r_{0}}} - (3 + 2 \alpha r_{0})  \right), \nonumber \\
&C_{1}& =  - 2(e^{2\alpha r_{0}} + 1) \left[ 3 \left( \frac{1 + e^{-2\alpha r_{0}}}{2\alpha r_{0}}\right) - (3 + 2 \alpha r_{0})  \left( \frac{1 + e^{-2\alpha r_{0}}}{2\alpha r_{0}}\right) \right], \nonumber \\
&C_{2}& =  (e^{2\alpha r_{0}} + 1)^{2} \left( \frac{1 + e^{-2\alpha r_{0}}}{2\alpha r_{0}}\right)^{2} \left[(3 + 2 \alpha r_{0})  -   \left( \frac{4\alpha r_{0}}{1 + e^{-2\alpha r_{0}}}\right) \right]
\label{t15},
\end{eqnarray}
other higher terms are neglected.

On using the transformation $z = -e^{-2\alpha r}$ and the Pekeris-type approximation in equation (\ref{t14}), equation (\ref{t13}) becomes 
\begin{equation}
z^{2}\frac{F_{n\kappa}^{2}(z)}{dz^{2}} + z\frac{F_{n\kappa}(z)}{dz} - \left[\nu^{2} + \frac{\eta V_{1}}{\alpha^{2}}\frac{z}{(1 - z)^{2}} + \frac{\zeta C_{2}}{4\alpha^{2}r_{0}^{2}}\frac{z^{2}}{(1 - z)^{2}} + \left(\frac{\eta V_{2}}{2\alpha^{2}} + \frac{\zeta C_{1}}{4\alpha^{2}r_{0}^{2}}  \right)\frac{z}{(1 - z)} \right]F_{n\kappa}(z) = 0  
\label{t16},
\end{equation}
where
\begin{eqnarray}
&\displaystyle{\Lambda_{\kappa} = \kappa + H + 1}, \nonumber \\
& \displaystyle{\eta = M + E_{n\kappa} - C_{s}},\nonumber \\
&\displaystyle{\beta_{1}^{2} = (M - E_{n\kappa})(M + E_{n\kappa} - C_{s})},\nonumber \\
&\displaystyle{\zeta = \Lambda_{\kappa} (\Lambda_{\kappa} - 1)}
\label{t17}.
\end{eqnarray}

By taking the function $F_{n\kappa}(r)$ to be
\begin{equation}  
F_{n\kappa}(z) = (1 - z)^{1 + \rho} z^{\nu}f_{n\kappa}(z)
\label{t18},
\end{equation}
in equation (\ref{t16}) and solve, the energy equation is obtained as
\begin{eqnarray}
&\displaystyle{(M + E_{n\kappa} - C_{s})(M - E_{n\kappa} + V_{2})   = - \frac{\zeta C_{0}}{r_{0}^{2}}} \nonumber \\
&\displaystyle{+ 4\alpha^{2}\left[ \frac{\frac{(C_{2} - C_{1})}{4\alpha^{2}r_{0}^{2}}\zeta - \frac{(M + E_{n\kappa} - C_{s})V_{2}}{2 \alpha^{2}}}{2(n + \rho + 1)}  - \frac{(n + \rho + 1)}{2}\right]^{2}} 
\label{t19}~,
\end{eqnarray}
where
\begin{equation}
\rho = \frac{1}{2} \left[-1 + \sqrt{1 + \frac{\zeta C_{2}}{\alpha^{2}r_{0}^{2}} + \frac{4 \eta V_{1}}{\alpha^{2}}} \right]
\label{t20},
\end{equation}
and $E_{n\kappa} \neq - M + C_{s}$, only positive energy solutions are valid.

The associated upper component spinor $F_{n\kappa}(r)$ is obtained as \cite{Oye12}
\begin{eqnarray}
&\displaystyle{F_{n\kappa}(r) = C_{n\kappa}(1 + e^{-2 \alpha r})^{1 + \rho} (- e^{-2\alpha r})^{\nu}~_{2}F_{1}(-n, n + 2(\nu + \rho + 1); 2\nu + 1; -e^{-2\alpha r})} \nonumber \\
& \displaystyle{ = C_{n\kappa}~\frac{n! \Gamma{(2\nu + 1)}}{\Gamma{(n + 2\nu + 1)}}(1 + e^{-2 \alpha r})^{1 + \rho} (- e^{-2\alpha r})^{\nu} P_{n}^{(2\nu, ~2\rho + 1)}(1 - 2z)}
\label{t21}~,
\end{eqnarray}
where
\begin{equation}
\nu =  \sqrt{\frac{\zeta C_{0}}{4 \alpha^{2} r_{0}^{2}} + \frac{\eta V_{2}}{4 \alpha^{2}} + \frac{\beta_{1}^{2}}{4 \alpha^{2}}}
\label{t22}
\end{equation}
and $C_{n\kappa}$ is the normalization constant which can easily be determined as \cite{Oye12}
\begin{equation}
\displaystyle{ C_{n \kappa} = \left[ \frac{ \Gamma{(2 \rho + 3)} \Gamma{(2 \nu + 1)} }{2 \alpha \Gamma{(n)}} \sum_{k = 0}^{\infty} \frac{(-1)^{k} \left(n + 2(1 + \nu + \rho)\right)_{k} \Gamma{(n + k)}}{k! (k + 2 \nu)!\Gamma{\left(k + 2(\nu + \rho + \frac{3}{2})\right)}} A_{n\kappa}  \right]^{-1/2}}
\label{t23}~,
\end{equation}
where $A_{n\kappa} =\,  _{3}F_{2}(2\nu + k, \, -n,\, n + 2(1 + \nu + \rho);\,k + 2(\nu + \rho + \frac{3}{2});\, 2 \nu + 1;\,1)$ and $(x)_{a} = \frac{\Gamma{(x + a)}}{\Gamma{(x)}}$ (Pochhammer symbol).

The lower component spinor of the Dirac equation for the Rosen-Morse potential plus a Coulomb-like tensor potential can be evaluated from the relation
\begin{equation}
G_{n\kappa}(r) = \frac{1}{M + E_{n\kappa} - C_{s}}\left[\frac{d}{dr} + \frac{\kappa}{r} - U(r) \right]F_{n\kappa}(r)
\label{t24}.
\end{equation}
In this case, it is noted that in the absence of the Coulomb-like-tensor interaction, the results above give the earlier results obtained for the Rosen-Morse potential discussed under the spin symmetry condition by Oyewumi (2012) \cite{Oye12}.

\subsection{Pseudopin Symmetric Bound State Solutions}
The exact pseudospin symmetry  has been discussed by Meng et al. (1998) and (1999) \cite{MeE98, MeE99}. This occurs in the Dirac equation when $\frac{d\Sigma(r)}{dr} = 0$ or $\Sigma(r) = C_{ps} = \mbox{constant}$. Again, a Coulomb-like potential as a tensor potential is added to the Rosen-Morse potential. Under this symmetry condition, the equation for the upper radial component spinor $G_{n\kappa}$ becomes
\begin{equation}
\left\{ \frac{d^{2}}{dr^{2}} - \frac{\kappa(\kappa - 1)}{r^{2}} - \frac{2\kappa}{r^{2}}H + \frac{H}{r^{2}} -  \frac{H^{2}}{r^{2}}  - \left[M + E_{n \kappa} - \Delta(r)\right] \left[M - E_{n \kappa} + C_{ps}\right] \right\} G_{n \kappa}(r)
\label{t25},
\end{equation}
with $ \kappa = \left\{
\begin{array}{ll}
\overline{\ell}, & \mbox{for}\; \kappa < 0  \\
- (\overline{\ell} + 1), & \mbox{for}\; \kappa >0~~,
\end{array}  \right. $ 
respectively. Equation (\ref{t25}) can be re-written with the substitution of the centrifugal term approximation scheme in equation (\ref{t14}) and on substituting $z = -e^{-2\alpha r}$, equation (\ref{t25}) becomes 
\begin{equation}
z^{2}\frac{G_{n\kappa}^{2}(z)}{dz^{2}} + z\frac{G_{n\kappa}(z)}{dz} - \left[\overline{\nu}^{2} + \frac{\overline{\eta} V_{1}}{\alpha^{2}}\frac{z}{(1 - z)^{2}} + \frac{\overline{\zeta} C_{2}}{4\alpha^{2}r_{0}^{2}}\frac{z^{2}}{(1 - z)^{2}} + \left(\frac{\overline{\eta} V_{2}}{2\alpha^{2}} + \frac{\overline{\zeta} C_{1}}{4\alpha^{2}r_{0}^{2}}  \right)\frac{z}{(1 - z)} \right]G_{n\kappa}(z) = 0 
\label{t26},
\end{equation}
where
\begin{eqnarray}
&\displaystyle{\overline{\Lambda}_{\kappa} = \kappa + H}, \nonumber \\
& \displaystyle{\overline{\eta} = E_{n\kappa} - M - C_{ps}},\nonumber \\
&\displaystyle{\overline{\beta}_{1}^{2} = (M + E_{n\kappa})(M - E_{n\kappa} + C_{ps})},\nonumber \\
&\displaystyle{\zeta = \overline{\Lambda}_{\kappa} (\overline{\Lambda}_{\kappa} - 1)}
\label{t27}.
\end{eqnarray}

Equation (\ref{t26}) is identical with the equation (\ref{t16}), therefore, the energy equation is obtained as
\begin{eqnarray}
&\displaystyle{(M - E_{n\kappa} +  C_{ps})(M + E_{n\kappa} - V_{2})   = - \frac{\overline{\zeta} C_{0}}{r_{0}^{2}}} \nonumber \\
&\displaystyle{+ 4\alpha^{2}\left[ \frac{\frac{(C_{2} - C_{1})}{4\alpha^{2}r_{0}^{2}}\overline{\zeta} - \frac{( E_{n\kappa} - M - C_{ps})V_{2}}{2 \alpha^{2}}}{2(n + \overline{\rho} + 1)}  - \frac{(n + \overline{\rho} + 1)}{2}\right]^{2}} 
\label{t28}~,
\end{eqnarray}
where
\begin{equation}
\rho = \frac{1}{2} \left[-1 + \sqrt{1 + \frac{\overline{\zeta} C_{2}}{\alpha^{2}r_{0}^{2}} + \frac{4 \overline{\eta} V_{1}}{\alpha^{2}}} \right]
\label{t29},
\end{equation}
and $E_{n\kappa} \neq M + C_{ps}$, only negative energy solutions are valid.

The corresponding upper component spinor $G_{n\kappa}(r)$ is obtained as
\begin{eqnarray}
&\displaystyle{G_{n\kappa}(r) = \overline{C}_{n\kappa}(1 + e^{-2 \alpha r})^{1 + \overline{\rho}} (- e^{-2\alpha r})^{\overline{\nu}}~_{2}F_{1}(-n, n + 2(\overline{\nu} + \overline{\rho} + 1); 2\overline{\nu} + 1; -e^{-2\alpha r})} \nonumber \\
& \displaystyle{ = \overline{C}_{n\kappa}~\frac{n! \Gamma{(2\overline{\nu} + 1)}}{\Gamma{(n + 2\overline{\nu} + 1)}}(1 + e^{-2 \alpha r})^{1 + \overline{\rho}} (- e^{-2\alpha r})^{\overline{\nu}} P_{n}^{(2\overline{\nu}, ~2\overline{\rho} + 1)}(1 - 2z)}
\label{t30}~,
\end{eqnarray}
where
\begin{equation}
\overline{\nu} =  \sqrt{\frac{\overline{\zeta} C_{0}}{4 \alpha^{2} r_{0}^{2}} + \frac{\overline{\eta} V_{2}}{4 \alpha^{2}} + \frac{\overline{\beta}_{1}^{2}}{4 \alpha^{2}}}
\label{t31}
\end{equation}
and $\overline{C}_{n\kappa}$ is the normalization constant is obtained \cite{Oye12}
\begin{equation}
\displaystyle{ \overline{C}_{n \kappa} = \left[ \frac{ \Gamma{(2 \overline{\rho} + 3)} \Gamma{(2 \overline{\nu} + 1)} }{2 \alpha \Gamma{(n)}} \sum_{k = 0}^{\infty} \frac{(-1)^{k} \left(n + 2(1 + \overline{\nu} + \overline{\rho})\right)_{k} \Gamma{(n + k)}}{k! (k + 2 \overline{\nu})!\Gamma{\left(k + 2(\overline{\nu} + \overline{\rho} + \frac{3}{2})\right)}} \overline{A}_{n\kappa}  \right]^{-1/2}}
\label{t32}~,
\end{equation}
where $\overline{A}_{n\kappa} =\,  _{3}F_{2}(2\overline{\nu} + k, \, -n,\, n + 2(1 + \overline{\nu} + \overline{\rho});\,k + 2(\overline{\nu} + \overline{\rho} + \frac{3}{2});\, 2 \overline{\nu} + 1;\,1)$ and $(x)_{a} = \frac{\Gamma{(x + a)}}{\Gamma{(x)}}$ (Pochhammer symbol).

The lower component spinor of the Dirac equation for the Rosen-Morse potential plus a Coulomb-like tensor potential can be evaluated from the relation
\begin{equation}
F_{n\kappa}(r) = \frac{1}{M - E_{n\kappa} + C_{ps}}\left[\frac{d}{dr} - \frac{\kappa}{r} + U(r) \right]G_{n\kappa}(r)
\label{t33}.
\end{equation}
Again, in this case, in the absence of the Coulomb-like-tensor interaction, the results above give the earlier results obtained for the Rosen-Morse potential discussed under the pseudospin symmetry condition by Oyewumi (2012) \cite{Oye12}.

\section{Solutions of a special case of the Dirac equation with  the Rosen-Morse potential (modified P\"{o}schl-Teller) including a tensor interaction}

It is very interesting to note that equation (\ref{t1}) reduces to the modified P\"{o}schl-Teller potential when $V_{2} = 0$  as \cite{EsM111, Agb11, OyA10, Oye12, OyE10, Gry+}:
\begin{equation}
V(r) = - V_{1} \textrm{sech}^{2}{\alpha r}
\label{t34},
\end{equation}
and when coupled with a tensor interaction, the PSS and SS solutions of the Dirac equation with modified P\"{o}schl-Teller including a Coulomb-like tensor potential are obtained.  Modified P\"{o}schl-Teller potential is a special case of the Rosen-Morse potential and is a typical $\Lambda$-nuclear potential which has been a very useful model in the study of $\Lambda$-hypernuclei in nuclear physics \cite{Gry+}.

For the SS solutions of the Modified P\"{o}schl-Teller and a Coulomb-like tensor potential, the energy equation is
\begin{equation}
(M + E_{n\kappa} - C_{s})(M - E_{n\kappa})   = - \frac{\zeta C_{0}}{r_{0}^{2}} 
+ 4\alpha^{2}\left[ \frac{\frac{(C_{2} - C_{1})}{4\alpha^{2}r_{0}^{2}}\zeta }{2(n + \rho + 1)}  - \frac{(n + \rho + 1)}{2}\right]^{2} 
\label{t35}
\end{equation}
and the associated upper component spinor $F_{n\kappa}(r)$ is 
\begin{eqnarray}
&\displaystyle{F_{n\kappa}(r) = N_{n\kappa}(1 + e^{-2 \alpha r})^{1 + \rho} (- e^{-2\alpha r})^{\nu_{1}}~_{2}F_{1}(-n, n + 2(\nu_{1} + \rho + 1); 2\nu_{1} + 1; -e^{-2\alpha r})} \nonumber \\
& \displaystyle{ = N_{n\kappa}~\frac{n! \Gamma{(2\nu_{1} + 1)}}{\Gamma{(n + 2\nu_{1} + 1)}}(1 + e^{-2 \alpha r})^{1 + \rho} (- e^{-2\alpha r})^{\nu_{1}} P_{n}^{(2\nu_{1}, ~2\rho + 1)}(1 - 2z)}
\label{t36}~,
\end{eqnarray}
where $ \rho = \frac{1}{2} \left[-1 + \sqrt{1 + \frac{\zeta C_{2}}{\alpha^{2}r_{0}^{2}} + \frac{4 \eta V_{1}}{\alpha^{2}}} \right]$, ~~$ \nu_{1} =  \sqrt{\frac{\zeta C_{0}}{4 \alpha^{2} r_{0}^{2}} + \frac{\beta_{1}^{2}}{4 \alpha^{2}}}$.
$N_{n\kappa}$ is the normalization constant obtained as \cite{Oye12}:
\begin{equation}
\displaystyle{ N_{n \kappa} = \left[ \frac{ \Gamma{(2 \rho + 3)} \Gamma{(2 \nu_{1} + 1)} }{2 \alpha \Gamma{(n)}} \sum_{k = 0}^{\infty} \frac{(-1)^{k} \left(n + 2(1 + \nu_{1} + \rho)\right)_{k} \Gamma{(n + k)}}{k! (k + 2 \nu_{1})!\Gamma{\left(k + 2(\nu_{1} + \rho + \frac{3}{2})\right)}} C_{n\kappa}  \right]^{-1/2}}
\label{t37}~,
\end{equation}
where $C_{n\kappa} =\,  _{3}F_{2}(2\nu_{1} + k, \, -n,\, n + 2(1 + \nu_{1} + \rho);\,k + 2(\nu_{1} + \rho + \frac{3}{2});\, 2 \nu_{1} + 1;\,1)$ and $(x)_{a} = \frac{\Gamma{(x + a)}}{\Gamma{(x)}}$ (Pochhammer symbol).  In a similar way as in Section $3.1$, the lower component spinor $G_{n\kappa}(r)$ of the Dirac equation for the P\"{o}schl-Teller potential plus a Coulomb-like tensor potential can be obtained. When the Coulomb-like tensor is removed, these results are identical with the results of Agboola (2011) for $N = 3$.

For the PSS solutions of the Modified P\"{o}schl-Teller and a Coulomb-like tensor potential, the energy equation is
\begin{equation}
(M - E_{n\kappa} +  C_{ps})(M + E_{n\kappa})   = - \frac{\overline{\zeta} C_{0}}{r_{0}^{2}} + 4\alpha^{2}\left[ \frac{\frac{(C_{2} - C_{1})}{4\alpha^{2}r_{0}^{2}}\overline{\zeta}}{2(n + \overline{\rho} + 1)}  - \frac{(n + \overline{\rho} + 1)}{2}\right]^{2} 
\label{t38}~,
\end{equation}
and the corresponding upper component spinor $G_{n\kappa}(r)$ is obtained as
\begin{eqnarray}
&\displaystyle{G_{n\kappa}(r) = \overline{N}_{n\kappa}(1 + e^{-2 \alpha r})^{1 + \overline{\rho}} (- e^{-2\alpha r})^{\overline{\nu}_{2}}~_{2}F_{1}(-n, n + 2(\overline{\nu}_{2} + \overline{\rho} + 1); 2\overline{\nu}_{2} + 1; -e^{-2\alpha r})} \nonumber \\
& \displaystyle{ = \overline{N}_{n\kappa}~\frac{n! \Gamma{(2\overline{\nu}_{2} + 1)}}{\Gamma{(n + 2\overline{\nu}_{2} + 1)}}(1 + e^{-2 \alpha r})^{1 + \overline{\rho}} (- e^{-2\alpha r})^{\overline{\nu}_{2}} P_{n}^{(2\overline{\nu}_{2}, ~2\overline{\rho} + 1)}(1 - 2z)}
\label{t39}~,
\end{eqnarray}
where $\rho = \frac{1}{2} \left[-1 + \sqrt{1 + \frac{\overline{\zeta} C_{2}}{\alpha^{2}r_{0}^{2}} + \frac{4 \overline{\eta} V_{1}}{\alpha^{2}}} \right]$, ~$ \overline{\nu}_{2} =  \sqrt{\frac{\overline{\zeta} C_{0}}{4 \alpha^{2} r_{0}^{2}} + \frac{\overline{\beta}_{1}^{2}}{4 \alpha^{2}}}$. The normalization constant $\overline{N}_{n\kappa}$ is obtained as \cite{Oye12}:
\begin{equation}
\displaystyle{ \overline{N}_{n \kappa} = \left[ \frac{ \Gamma{(2 \overline{\rho} + 3)} \Gamma{(2 \overline{\nu}_{2} + 1)} }{2 \alpha \Gamma{(n)}} \sum_{k = 0}^{\infty} \frac{(-1)^{k} \left(n + 2(1 + \overline{\nu}_{2} + \overline{\rho})\right)_{k} \Gamma{(n + k)}}{k! (k + 2 \overline{\nu}_{2})!\Gamma{\left(k + 2(\overline{\nu}_{2} + \overline{\rho} + \frac{3}{2})\right)}} \overline{A}_{n\kappa}  \right]^{-1/2}}
\label{t40}~,
\end{equation}
where $\overline{A}_{n\kappa} =\,  _{3}F_{2}(2\overline{\nu}_{2} + k, \, -n,\, n + 2(1 + \overline{\nu}_{2} + \overline{\rho});\,k + 2(\overline{\nu}_{2} + \overline{\rho} + \frac{3}{2});\, 2 \overline{\nu}_{2} + 1;\,1)$ and $(x)_{a} = \frac{\Gamma{(x + a)}}{\Gamma{(x)}}$ (Pochhammer symbol). 

The results obtained in this Section are identical with the results of Eshghi and Mehraban (2011) for $q = 1$.

\section{Conclusions}

It has been shown that the approximate solutions of the Dirac equation with equal scalar and vector Rosen-Morse potentials and a Coulomb-like tensor potential can be obtained by using the Pekeris-type approximation scheme to deal with the (pseudo or) centrifugal term. We have used the following parameters in the numerical calculations: $r_{0} = 2.40873 fm,  V_{1} = 0.001 fm^{-1}, V_{2} = 0.01 fm^{-1}, C_{0} = 0.26928, C_{1} = 0.62178, C_{3} = 0.10893, \alpha = 0.988879, M = 10 fm^{-1}$.

 Table $1$ \& $2$ contain the bound state energy eigenvalues ($fm^{-1}$) of the spin symmetry Rosen-Morse potential with $H = 0.0,~ 0.5,~ 1.0$ for some values of $n$ and $\kappa$  when $C_{s} = 10 fm^{-1}$ and $C_{s} = 0$, respectively. Table $3$ contains the bound state energy eigenvalues ($fm^{-1}$) of the spin symmetry modified P\"{o}schl-Teller potential with $H = 0.0,~ 0.5,~ 1.0$ for some values of $n$ and $\kappa$  when $C_{s} = 10 fm^{-1}$.

In the same way, Table $4$ \& $5$ contain the bound state energy eigenvalues ($fm^{-1}$) of the pseudospin symmetry Rosen-Morse potential with $H = 0.0,~ 0.5,~ 1.0$ for some values of $n$ and $\kappa$  when $C_{s} = 10 fm^{-1}$ and $C_{s} = 0$, respectively. Table $6$ contains the bound state energy eigenvalues ($fm^{-1}$) of the pseudospin symmetry modified P\"{o}schl-Teller potential with $H = 0.0,~ 0.5,~ 1.0$ for some values of $n$ and $\kappa$  when $C_{s} = 10 fm^{-1}$.

The numerical results obtained in Table 1 - 3 show that, in the spin symmetry limit, the degenerate states for various $H$ are as follows:

\noindent
For ${\bf H = 0;}$
  
\noindent
$0p_{\frac{3}{2}} = 0p_{\frac{1}{2}}, 1p_{\frac{3}{2}} = 1p_{\frac{1}{2}}, 2p_{\frac{3}{2}} = 2p_{\frac{1}{2}}, 3p_{\frac{3}{2}} = 3p_{\frac{1}{2}}, 0d_{\frac{5}{2}} = 0d_{\frac{3}{2}}, 1d_{\frac{5}{2}} = 1d_{\frac{3}{2}}, 2d_{\frac{5}{2}} = 2d_{\frac{3}{2}}, 3d_{\frac{5}{2}} = 3d_{\frac{3}{2}}, 0f_{\frac{7}{2}} = 0f_{\frac{5}{2}}, 1f_{\frac{7}{2}} = 1f_{\frac{5}{2}}, 2f_{\frac{7}{2}} = 2f_{\frac{5}{2}}, 3f_{\frac{7}{2}} = 3f_{\frac{5}{2}}$

\noindent
For ${\bf H = 0.5;}$
  
\noindent
$0d_{\frac{5}{2}} = 0p_{\frac{1}{2}}, 1d_{\frac{5}{2}} = 1p_{\frac{1}{2}}, 2d_{\frac{5}{2}} = 2p_{\frac{1}{2}}, 3d_{\frac{5}{2}} = 3p_{\frac{1}{2}}, 0f_{\frac{7}{2}} = 0d_{\frac{3}{2}}, 1f_{\frac{7}{2}} = 1d_{\frac{3}{2}}, 2f_{\frac{7}{2}} = 2d_{\frac{3}{2}}, 3f_{\frac{7}{2}} = 3d_{\frac{3}{2}}$

\noindent
For ${\bf H = 1.0;}$
  
\noindent
$0s_{\frac{1}{2}} = 0p_{\frac{3}{2}}, 1s_{\frac{1}{2}} = 1p_{\frac{3}{2}}, 2s_{\frac{1}{2}} = 2p_{\frac{3}{2}}, 3s_{\frac{1}{2}} = 3p_{\frac{3}{2}}, 0f_{\frac{7}{2}} = 0p_{\frac{1}{2}}, 1f_{\frac{7}{2}} = 1p_{\frac{1}{2}}, 2f_{\frac{7}{2}} = 2p_{\frac{1}{2}}, 3f_{\frac{7}{2}} = 3p_{\frac{1}{2}}$.

The degenerate states in the pseudospin symmetry limit for various $H$ as  shown in Table 4 - 6 are as follows:

\noindent
For ${\bf H = 0;}$
  
\noindent
$1s_{\frac{1}{2}} = 0d_{\frac{3}{2}}, 1p_{\frac{3}{2}} = 0f_{\frac{5}{2}}, 1d_{\frac{5}{2}} = 0g_{\frac{7}{2}}, 1f_{\frac{7}{2}} = 0h_{\frac{9}{2}}, 2s_{\frac{1}{2}} = 1d_{\frac{3}{2}}, 2p_{\frac{3}{2}} = 1f_{\frac{5}{2}}, 2d_{\frac{5}{2}} = 1g_{\frac{7}{2}}, 2f_{\frac{7}{2}} = 1h_{\frac{9}{2}}$

\noindent
For ${\bf H = 0.5;}$
  
\noindent
$1pd_{\frac{3}{2}} = 0d_{\frac{3}{2}}, 1d_{\frac{5}{2}} = 0f_{\frac{5}{2}}, 1f_{\frac{7}{2}} = 0g_{\frac{7}{2}}, 2p_{\frac{3}{2}} = 1d_{\frac{3}{2}}, 2d_{\frac{5}{2}} = 1f_{\frac{5}{2}}, 2f_{\frac{7}{2}} = 1g_{\frac{7}{2}}$

\noindent
For ${\bf H = 1.0;}$
  
\noindent
$1d_{\frac{5}{2}} = 0d_{\frac{3}{2}}, 1f_{\frac{7}{2}} = 0f_{\frac{5}{2}}, 2d_{\frac{5}{2}} = 1d_{\frac{3}{2}}, 2f_{\frac{7}{2}} = 1f_{\frac{5}{2}}$

From the numerical results, it has also been shown that the degeneracy between spin doublets and pseudo-spin doublets is removed by tensor interaction for the two potentials considered.

\vspace{15mm}
{\bf  \large{Acknowledgments}.}

\vspace{5mm}
{ \footnotesize
The author thanks his host Prof. K. D. Sen of the School of Chemistry, University of Hyderabad, India during his TWAS-UNESCO Associate research visit where part of this work has been done. Also, he thanks his host Prof. M. N. Hounkonnou the President of the ICMPA-UNESCO Chair, University of Abomey-Calavi, Republic of Benin where this work has been finalized. He acknowledges the University of Ilorin for granting him leave. eJDS (ICTP) is acknowledged. 
 Also, he appreciates the efforts of Profs. Ginocchio, J. N., Hasanabadi, H.,  Drs. Ta\c{s}kin, F., Hamzavi, M., Aydo$\check{g}$du, O. Eshighi, M. and  Maghsoodi, E. for their assistance.
}
 \\

\begin{table}[!hbp]
\caption{The bound state energy eigenvalues ($fm^{-1}$) of the spin symmetry Rosen-Morse potential for some values of $n$ and $\kappa$ with $H = 0.0, 0.5 ~\mbox{and}~ 1.0$ ($C_{s} = 10 fm^{-1}$).  \vspace*{13pt}} {\footnotesize
\begin{tabular}{|c c c c c c c|}\hline
{}&{} &{} &{}&{}&{}&{} \\[-1.5ex]
$n$& $\ell$ &$\kappa$ & $(\ell, j)$ & $E_{n, \kappa}$ (H = 0.0)& $E_{n, \kappa}$ (H = 0.5)& $E_{n, \kappa}$ (H = 1.0) \\[2ex]\hline
0&	 0& 	  -1&   $0s_{1/2}$&    9.89898&   9.89921&  9.89898\\[1ex]
1&	 0&   	-1&	  $1s_{1/2}$&    9.58799&	  9.58846&  9.58799\\[1ex]
2&	 0&    	-1&	  $2s_{1/2}$&    9.01813&  	9.01890&  9.01813\\[1ex]
3&   0&	    -1& 	$3s_{1/2}$&    8.04795& 	8.04936&  8.04795\\[1ex]
0&	 1&	    -2&	  $0p_{3/2}$&    9.89691&	  9.89825&  9.89898\\[1ex]
1&   1&   	-2& 	$1p_{3/2}$&    9.58432& 	9.58663&  9.58799\\[1ex]
2&   1&	    -2&	  $2p_{3/2}$&    9.01172& 	9.01571&  9.01813\\[1ex]
3&   1&   	-2& 	$3p_{3/2}$&    8.03655& 	8.04365&  8.04795\\[1ex]
0&	 2&	    -3&	  $0d_{5/2}$&    9.89174&	  9.89482&  9.89691\\[1ex]
1&	 2&	    -3&	  $1d_{5/2}$&    9.57677&	  9.58104&  9.58432\\[1ex]
2&	 2&	    -3&	  $2d_{5/2}$&    8.99906&	  9.00617&  9.01172\\[1ex]
3&	 2&	    -3&	  $3d_{5/2}$&    8.01407& 	8.02667&  8.03655\\[1ex]
0&	 3&	    -4&	  $0f_{7/2}$&    9.88168&	  9.88746&  9.89174\\[1ex]
1&	 3&	    -4&	  $1f_{7/2}$&    9.56506&	  9.57146&  9.57677\\[1ex]
2&	 3&	    -4&	  $2f_{7/2}$&    8.98028& 	8.99042&  8.99906\\[1ex]
3& 	 3&	    -4&	  $3f_{7/2}$&    7.98103&	  7.99884&  8.01407\\[1ex]
0&	 1&	     1&	  $0p_{1/2}$&    9.89691&	  9.89482&  9.89174\\[1ex]
1&	 1&	     1&	  $1p_{1/2}$&    9.58432&	  9.58104&  9.57677\\[1ex]
2&	 1&	     1&	  $2p_{1/2}$&    9.01172&	  9.00617&  8.99906\\[1ex]
3&	 1&	     1&	  $3p_{1/2}$&    8.03655& 	8.02667&  8.01407\\[1ex]
0&	 2&	     2&	  $0d_{3/2}$&    9.89174&	  9.88746&  9.88168\\[1ex]
1&	 2&	     2&	  $1d_{3/2}$&    9.57677&   9.57146&	9.56506\\[1ex]
2&	 2&	     2&	  $2d_{3/2}$&    8.99906&	  8.99042&	8.98028\\[1ex]
3&	 2&	     2&	  $3d_{3/2}$&    8.01407&	  7.99884&  7.98103\\[1ex]
0&	 3&	     3&	  $0f_{5/2}$&    9.88168&   9.87411&	9.86445\\[1ex]
1&	 3&	     3&	  $1f_{5/2}$&    9.56506&	  9.55749&	9.54869\\[1ex]
2&	 3&	     3&	  $2f_{5/2}$&    8.98028&	  8.96863&	8.95552\\[1ex]
3&	 3&	     3&	  $3f_{5/2}$&    7.98103& 	7.96072&	7.93793\\[1ex]\hline
\end{tabular}\label{tab1} }
\vspace*{-13pt}
\end{table}

\begin{table}[!hbp]
\caption{The bound state energy eigenvalues ($fm^{-1}$) of the spin symmetry Rosen-Morse potential for some values of $n$ and $\kappa$ with $H = 0.0, 0.5 ~\mbox{and}~ 1.0$ ($C_{s} = 0$). \vspace*{13pt}} {\footnotesize
\begin{tabular}{|c c c c c c c|}\hline
{}&{} &{} &{}&{}&{}&{} \\[-1.5ex]
$n$& $\ell$ &$\kappa$ & $(\ell, j)$ & $E_{n, \kappa}$ (H = 0.0)& $E_{n, \kappa}$ (H = 0.5)& $E_{n, \kappa}$ (H = 1.0) \\[2ex]\hline
0&	 0& 	  -1&   $0s_{1/2}$&    9.94851& 	9.94865&  9.94851\\[1ex]
1&	 0&   	-1&	  $1s_{1/2}$&    9.79836&  	9.79858&  9.79836\\[1ex]
2&	 0&    	-1&	  $2s_{1/2}$&    9.54372&   9.54405&  9.54372\\[1ex]
3&   0&	    -1& 	$3s_{1/2}$&    9.17619& 	9.17665&  9.17619\\[1ex]
0&	 1&	    -2&	  $0p_{3/2}$&    9.94730&   9.94807&  9.94851\\[1ex]
1&   1&   	-2& 	$1p_{3/2}$&    9.79661&   9.79771&  9.79836\\[1ex]
2&   1&	    -2&	  $2p_{3/2}$&    9.54106& 	9.54271&  9.54372\\[1ex]
3&   1&   	-2& 	$3p_{3/2}$&    9.17246&   9.17479&  9.17619\\[1ex]
0&	 2&	    -3&	  $0d_{5/2}$&    9.94439&   9.94611&  9.94730\\[1ex]
1&	 2&	    -3&	  $1d_{5/2}$&    9.79302&   9.79505&  9.79661\\[1ex]
2&	 2&	    -3&	  $2d_{5/2}$&    9.53577&   9.53873&  9.54106\\[1ex]
3&	 2&	    -3&	  $3d_{5/2}$&    9.16514&   9.16925&  9.17246\\[1ex]
0&	 3&	    -4&	  $0f_{7/2}$&    9.93897&   9.94206&  9.94439\\[1ex]
1&	 3&	    -4&	  $1f_{7/2}$&    9.78747&   9.79050&  9.79302\\[1ex]
2&	 3&	    -4&	  $2f_{7/2}$&    9.52796&   9.53218&  9.53577\\[1ex]
3& 	 3&	    -4&	  $3f_{7/2}$&    9.15446&   9.16021&  9.16514\\[1ex]
0&	 1&	     1&	  $0p_{1/2}$&    9.94730&   9.94611&  9.94439\\[1ex]
1&	 1&	     1&	  $1p_{1/2}$&    9.79661&   9.79505&  9.79302\\[1ex]
2&	 1&	     1&	  $2p_{1/2}$&    9.54106&   9.53873&  9.53577\\[1ex]
3&	 1&	     1&	  $3p_{1/2}$&    9.17246&   9.16925&  9.16514\\[1ex]
0&	 2&	     2&	  $0d_{3/2}$&    9.94439&   9.94206&  9.93897\\[1ex]
1&	 2&	     2&	  $1d_{3/2}$&    9.79302&   9.79050&  9.78747\\[1ex]
2&	 2&	     2&	  $2d_{3/2}$&    9.53577&   9.53218&  9.52796\\[1ex]
3&	 2&	     2&	  $3d_{3/2}$&    9.16514&   9.16021&  9.15446\\[1ex]
0&	 3&	     3&	  $0f_{5/2}$&    9.93897&   9.93496&  9.92999\\[1ex]
1&	 3&	     3&	  $1f_{5/2}$&    9.78747&   9.78390& 	9.77973\\[1ex]
2&	 3&	     3&	  $2f_{5/2}$&    9.52796&   9.52312&	9.51767\\[1ex]
3&	 3&	     3&	  $3f_{5/2}$&    9.16186&   9.14794&  9.14065\\[1ex]\hline
\end{tabular}\label{tab2} }
\vspace*{-13pt}
\end{table}

\begin{table}[!hbp]
\caption{The bound state energy eigenvalues ($fm^{-1}$) of the spin symmetry P\"{o}schl-Teller potential for some values of $n$ and $\kappa$ with $H = 0.0, 0.5 ~\mbox{and}~ 1.0$ ($C_{s} = 10 fm^{-1}, V_{2} = 0 ~\mbox{in equation (\ref{t1}), i. e. Rosen-Morse potential}).$  \vspace*{13pt}} {\footnotesize
\begin{tabular}{|c c c c c c c|}\hline
{}&{} &{} &{}&{}&{}&{} \\[-1.5ex]
$n$& $\ell$ &$\kappa$ & $(\ell, j)$ & $E_{n, \kappa}$ (H = 0.0)& $E_{n, \kappa}$ (H = 0.5)& $E_{n, \kappa}$ (H = 1.0) \\[2ex]\hline
0&	 0& 	  -1&   $0s_{1/2}$&    9.89923&	  9.89940&  9.89923\\[1ex]
1&	 0&   	-1&	  $1s_{1/2}$&    9.58807&	  9.58851&  9.58807\\[1ex]
2&	 0&    	-1&	  $2s_{1/2}$&    9.01813&	  9.01892&  9.01813\\[1ex]
3&   0&	    -1& 	$3s_{1/2}$&    8.04794&	  8.04937&  8.04794\\[1ex]
0&	 1&	    -2&	  $0p_{3/2}$&    9.89760&	  9.89866&  9.89923\\[1ex]
1&   1&   	-2& 	$1p_{3/2}$&    9.58450&	  9.58674&  9.58807\\[1ex]
2&   1&	    -2&	  $2p_{3/2}$&    9.01181&	  9.01576&  9.01813\\[1ex]
3&   1&   	-2& 	$3p_{3/2}$&    8.03659&	  8.04369&  8.04794\\[1ex]
0&	 2&	    -3&	  $0d_{5/2}$&    9.89326&	  9.89587&  9.89760\\[1ex]
1&	 2&	    -3&	  $1d_{5/2}$&    9.57717& 	9.58132&  9.58450\\[1ex]
2&	 2&	    -3&	  $2d_{5/2}$&    8.99924&	  9.00629&  9.01181\\[1ex]
3&	 2&	    -3&	  $3d_{5/2}$&    8.01420&	  8.02675&  8.03659\\[1ex]
0&	 3&	    -4&	  $0f_{7/2}$&    9.88433&	  9.88951&  9.89326\\[1ex]
1&	 3&	    -4&	  $1f_{7/2}$&    9.56579&	  9.57202&  9.57717\\[1ex]
2&	 3&	    -4&	  $2f_{7/2}$&    8.98064&	  8.99069&  8.99924\\[1ex]
3& 	 3&	    -4&	  $3f_{7/2}$&    7.98126&	  7.99901&  8.01420\\[1ex]
0&	 1&	     1&	  $0p_{1/2}$&    9.89760&	  9.89587&  9.89326\\[1ex]
1&	 1&	     1&	  $1p_{1/2}$&    9.58450&	  9.58132&  9.57717\\[1ex]
2&	 1&	     1&	  $2p_{1/2}$&    9.01181&	  9.00629&  8.99924\\[1ex]
3&	 1&	     1&	  $3p_{1/2}$&    8.03659&	  8.02675&  8.01420\\[1ex]
0&	 2&	     2&	  $0d_{3/2}$&    9.89326& 	9.88951&  9.88433\\[1ex]
1&	 2&	     2&	  $1d_{3/2}$&    9.57717&	  9.57202&  9.56579\\[1ex]
2&	 2&	     2&	  $2d_{3/2}$&    8.99924&  	8.99069&  8.98064\\[1ex]
3&	 2&	     2&	  $3d_{3/2}$&    8.01420&   7.99901&  7.98126\\[1ex]
0&	 3&	     3&	  $0f_{5/2}$&    9.88433&	  9.87742&  9.86847\\[1ex]
1&	 3&	     3&	  $1f_{5/2}$&    9.56579&	  9.55841&  9.54984\\[1ex]
2&	 3&	     3&	  $2f_{5/2}$&    8.98064&	  8.96909&  8.95608\\[1ex]
3&	 3&	     3&	  $3f_{5/2}$&    7.98126&	  7.96100&  7.93832\\[1ex]\hline
\end{tabular}\label{tab3} }
\vspace*{-13pt}
\end{table}

\begin{table}[!hbp]
\caption{The bound state energy eigenvalues ($fm^{-1}$) of the pseudospin symmetry Rosen-Morse potential for some values of $n$ and $\kappa$ with $H = 0.0, 0.5 ~\mbox{and}~ 1.0$ ($C_{ps} = -10 fm^{-1}$).   \vspace*{13pt}} {\footnotesize
\begin{tabular}{|c c c c c c c|}\hline
{}&{} &{} &{}&{}&{}&{} \\[-1.5ex]
$n$& $\overline{\ell}$ &$\kappa$ & $(\ell, j)$ & $E_{n, \kappa}$ (H = 0.0)& $E_{n, \kappa}$ (H = 0.5)& $E_{n, \kappa}$ (H = 1.0)  \\[2ex]\hline
1&	1&	-1& 	$1s_{1/2}$&  -9.58455&  	-9.58674&  -9.58799\\[1ex]
1&	2&	-2&	  $1p_{3/2}$&  -9.57745&  	-9.58147&  -9.58455\\[1ex]
1&	3&	-3&	  $1d_{5/2}$&  -9.56640&    -9.57245&  -9.57745\\[1ex]
1&	4&	-4&	  $1f_{7/2}$&  -9.55085&    -9.55921&  -9.56640\\[1ex]
2&	1&	-1&	  $2s_{1/2}$&  -9.01183&  	-9.01576&  -9.01813\\[1ex]
2&	2&	-2&	  $2p_{3/2}$&  -8.99939&    -9.00637&  -9.01183\\[1ex]
2&	3&	-3&	  $2d_{5/2}$&  -8.98094&  	-8.99090&  -8.99939\\[1ex]
2&	4&	-4&	  $2f_{7/2}$&  -8.95659&  	-8.96949&  -8.98094\\[1ex]
1&	1&	 2&	  $0d_{3/2}$&  -9.58455&  	-9.58147&  -9.57745\\[1ex]
1&	2&	 3&	  $0f_{5/2}$&  -9.57745&  	-9.57245&  -9.56640\\[1ex]
1&	3&	 4&	  $0g_{7/2}$&  -9.56640&  	-9.55921&  -9.55080\\[1ex]
1&	4&	 5&	  $0h_{9/2}$&  -9.55085&  	-9.54120&  -9.53018\\[1ex]
2&	1&	 2& 	$1d_{3/2}$&  -9.01183& 	  -9.00637&  -8.99939\\[1ex]
2&	2&	 3&	  $1f_{5/2}$&  -8.99939& 	  -8.99090&  -8.98094\\[1ex]
2&	3&	 4& 	$1g_{7/2}$&  -8.98094&  	-8.96949&  -8.95659\\[1ex]
2&	4&	 5& 	$1h_{9/2}$&  -8.95659& 	  -8.94222&  -8.92637\\[1ex]\hline
\end{tabular}\label{tab4} }
\vspace*{-13pt}
\end{table}

\begin{table}[!hbp]
\caption{The bound state energy eigenvalues ($fm^{-1}$) of the pseudospin symmetry Rosen-Morse potential for some values of $n$ and $\kappa$ with $H = 0.0, 0.5 ~\mbox{and}~ 1.0$ ($C_{ps} = 0$). \vspace*{13pt}} {\footnotesize
\begin{tabular}{|c c c c c c c|}\hline
{}&{} &{} &{}&{}&{}&{} \\[-1.5ex]
$n$& $\overline{\ell}$ &$\kappa$ & $(\ell, j)$ & $E_{n, \kappa}$ (H = 0.0)& $E_{n, \kappa}$ (H = 0.5)& $E_{n, \kappa}$ (H = 1.0)  \\[2ex]\hline
1&	1&	-1& 	$1s_{1/2}$&  -9.79683&    -9.79780&	 -9.79836\\[1ex]
1&	2&	-2&	  $1p_{3/2}$&  -9.79367&    -9.79546&	 -9.79683\\[1ex]
1&	3&	-3&	  $1d_{5/2}$&  -9.78874&    -9.79144&  -9.79367\\[1ex]
1&	4&	-4&	  $1f_{7/2}$&  -9.78179&    -9.78554&  -9.78874\\[1ex]
2&	1&	-1&	  $2s_{1/2}$&  -9.54116&    -9.54276&  -9.54372\\[1ex]
2&	2&	-2&	  $2p_{3/2}$&  -9.53607&    -9.53892&  -9.54116\\[1ex]
2&	3&	-3&	  $2d_{5/2}$&  -9.52854&    -9.53261&  -9.53607\\[1ex]
2&	4&	-4&	  $2f_{7/2}$&  -9.51863&    -9.52387&  -9.52854\\[1ex]
1&	1&	 2&	  $0d_{3/2}$&  -9.79683&    -9.79546&  -9.79368\\[1ex]
1&	2&	 3&	  $0f_{5/2}$&  -9.79368&    -9.79145&  -9.78874\\[1ex]
1&	3&	 4&	  $0g_{7/2}$&  -9.78874&    -9.78554&  -9.78179\\[1ex]
1&	4&	 5&	  $0h_{9/2}$&  -9.78179&    -9.77747&  -9.77253\\[1ex]
2&	1&	 2& 	$1d_{3/2}$&  -9.54116&    -9.53892&  -9.53607\\[1ex]
2&	2&	 3&	  $1f_{5/2}$&  -9.53607&    -9.53261&  -9.52854\\[1ex]
2&	3&	 4& 	$1g_{7/2}$&  -9.52854&    -9.52388&  -9.51863\\[1ex]
2&	4&	 5& 	$1h_{9/2}$&  -9.51863&    -9.51279&	 -9.50637\\[1ex]\hline
\end{tabular}\label{tab5} }
\vspace*{-13pt}
\end{table}

\begin{table}[!hbp]
\caption{The bound state energy eigenvalues ($fm^{-1}$) of the pseudospin symmetry P\"{o}schl-Teller potential for some values of $n$ and $\kappa$ with $H = 0.0, 0.5 ~\mbox{and}~ 1.0$ ($C_{ps} = -10 fm^{-1}, V_{2} = 0 ~\mbox{in equation (\ref{t1}), i. e. Rosen-Morse potential}).$  \vspace*{13pt}} {\footnotesize
\begin{tabular}{|c c c c c c c|}\hline
{}&{} &{} &{}&{}&{}&{} \\[-1.5ex]
$n$& $\overline{\ell}$ &$\kappa$ & $(\ell, j)$ & $E_{n, \kappa}$ (H = 0.0)& $E_{n, \kappa}$ (H = 0.5)& $E_{n, \kappa}$ (H = 1.0)  \\[2ex]\hline
1&	1&	-1& 	$1s_{1/2}$&  -9.58450&   	-9.58674&  -9.58807\\[1ex]
1&	2&	-2&	  $1p_{3/2}$&  -9.57718&  	-9.58132&  -9.58450\\[1ex]
1&	3&	-3&	  $1d_{5/2}$&  -9.56579&   	-9.57202&  -9.57718\\[1ex]
1&	4&	-4&	  $1f_{7/2}$&  -9.54983&  	-9.55841&  -9.56579\\[1ex]
2&	1&	-1&	  $2s_{1/2}$&  -9.01180&  	-9.01575&  -9.01813\\[1ex]
2&	2&	-2&	  $2p_{3/2}$&  -8.99925&  	-9.00630&  -9.01180\\[1ex]
2&	3&	-3&	  $2d_{5/2}$&  -8.98064&  	-8.99069&  -8.99925\\[1ex]
2&	4&	-4&	  $2f_{7/2}$&  -8.95608&    -8.96910&  -8.98064\\[1ex]
1&	1&	 2&	  $0d_{3/2}$&  -9.58450&    -9.58132&  -9.57718\\[1ex]
1&	2&	 3&	  $0f_{5/2}$&  -9.57718&  	-9.57201&  -9.56579\\[1ex]
1&	3&	 4&	  $0g_{7/2}$&  -9.56579&   	-9.55841&  -9.54983\\[1ex]
1&	4&	 5&	  $0h_{9/2}$&  -9.54983&  	-9.53994&  -9.52867\\[1ex]
2&	1&	 2& 	$1d_{3/2}$&  -9.01180&  	-9.00630&  -8.99925\\[1ex]
2&	2&	 3&	  $1f_{5/2}$&  -8.99925&    -8.99069&  -8.98064\\[1ex]
2&	3&	 4& 	$1g_{7/2}$&  -8.98064&  	-8.96910&  -8.95608\\[1ex]
2&	4&	 5& 	$1h_{9/2}$&  -8.95608&    -8.94159&  -8.92561\\[1ex]\hline
\end{tabular}\label{tab6} }
\vspace*{-13pt}
\end{table}

\end{document}